\documentstyle{article}
\begin{document}

\title{{\bf The N-steps Invasion Percolation Model}} 
\author{Reginaldo A. Zara\thanks{razara@ifsc.sc.usp.br} and Roberto N.
Onody\thanks{onody@ifsc.sc.usp.br}\\ \\ 
{\small {\em Departamento de F\'{\i}sica e Inform\'{a}tica } }\\ 
{\small {\em Instituto de F\'{\i}sica de S\~{a}o Carlos} }\\ 
{\small {\em Universidade de S\~{a}o Paulo - Caixa Postal 369} }\\ 
{\small {\em 13560-970 - S\~{a}o Carlos, S\~{a}o Paulo, Brasil.}}}
\date{} 
\maketitle 
\normalsize 
\baselineskip=16pt 

\begin{abstract} 

A new kind of invasion percolation is introduced in order to take into account
the inertia of the invader fluid. The inertia strength is controlled by the
number $N$ of pores (or steps) invaded after the perimeter rupture.
The new model belongs to a different class of universality with the fractal
dimensions of the percolating clusters depending on $N$. 
A blocking phenomenon takes place in two
dimensions. It imposes an upper bound value on $N$. For pore sizes larger than
the critical threshold, the acceptance profile exhibits a permanent tail.

\vspace{3cm}



\end{abstract}

\newpage

\section{\bf Introduction}

When a nonviscous liquid is slowly injected into a porous medium already filled
with a viscous fluid the predominant forces on the interface are of capillary
nature.  These forces are such as to make the injected fluid spontaneously
displace the viscous one.  The interface between the fluids advances pore by
pore, being its dynamics determined by the capillary rule:  the smaller pore is
invaded first.

Invasion percolation \cite{wil} is a theoretical model used to describe
fluid-fluid displacement.  As it was pointed out invasion
percolation is a kind of self-organizing criticality \cite{bak,furu}.  The
system exhibits scale invariant behaviour in space and time achieving the
critical state without a fine tuning mechanism to a particular parameter.
There is now a strong evidence that this critical state corresponds to the
critical standard percolation \cite{wil,stau}.

Under some modifications, the invasion percolation model has been successfully
applied to describe the fingering phenomena in soils \cite{ono1} and fluid
flowing with a privileged direction \cite{ono2}.  In its original formulation,
the invasion percolation assumes that all the pores situated in the perimeter
of the cluster exchange informations in such a way that, at growth stage $t$, no
matter how far they are, only an unique pore is invaded - that one with the
smallest size.  After this pore has been invaded the fluid flow stops, a new
perimeter is determined and the process continues.  Recently, a multiple
invasion percolation model \cite{ono3,ono4} was proposed permitting that not 
just
one but many pores belonging to the actual perimeter can be simultaneously
invaded.  But here again, the fluid does not exhibit inertia.  By this we mean
the tendency of the fluid to proceed further, invading the sites surrounding
that pore of the perimeter where the invasion process is taking place.  
We implement this idea by proposing a
model in which, at the growth stage $t$, the fluid
occupies not only the smaller perimeter site $j$ but also invades (one by one
and always following the capillary rule) $N-1$ additional pores on $j$'s 
neighborhood.
At $t+1$, a new perimeter is determined and the process continues.  We can say
that the fluid walks (occupies) $N$ steps (pores) before to loose its inertia.
We call this process the N-steps invasion percolation.  The ordinary invasion
percolation corresponds to the case $N=1$.

The N-steps invasion percolation model exhibits very different behaviours in two
and three dimensions.  In two dimensions and for large $N$, the walks can be 
easily blocked since hindrances to the growth are set everywhere by parts of the 
own cluster. The walks are most of the time incomplete.
In such walks, the number of steps actually executed is always equal or
smaller than the external (and fixed) parameter $N$. The mean number of steps
$ \langle n \rangle $
is bounded and its maximum value depends on the lattice geometry.
For an infinite square lattice we find that $ \langle n \rangle $ cannot be 
greater than 35.
The clusters 
have fractal dimensions very close to that of the ordinary invasion percolation
but very different mean coordination numbers. The acceptance profiles show
a strong dependence of the critical threshold with $ \langle n \rangle $.

In three dimensions, the possibility of blocking is so small that 
$ \langle n \rangle $ always coincides with $N$. 
The difference between the two and three dimensional behaviours 
is reminiscent of what happens in many growth models like, e. g., 
in the invasion percolation model with trapping \cite{wil}.
The calculated fractal 
dimensions obtained from a careful finite size scaling analysis, indicate 
that the N-steps model and the ordinary invasion percolation belong to different
classes of universality.

\section{\bf The Model}

Before treating the N-steps invasion percolation model,
let us briefly recall how the ordinary invasion percolation is simulated. 
Assign a random number $r$, uniformly distributed in the range
$[0,1]$ to each lattice site and choose the central site as the seed 
of the growth. The perimeter sites of the cluster are identified as the growth
sites. At each growth step, occupy an unique perimeter site - that one 
with the smallest associated random number.  The growth process is interrupted 
after the cluster reaches the lattice boundary. It is important to note that
in the ordinary invasion percolation the occupation of {\it one site}
corresponds to {\it one growth step}.  

In the construction of the N-steps invasion percolation cluster we define 
{\it one growth step}, taking place at the growth stage $t$ (which shall not 
be confused with time, see below for details), as being composed by the 
following procedures. First, as in the
ordinary invasion percolation, the smallest perimeter site is occupied. Then, 
starting from {\em this} site, additional $N-1$ sites are sequentially invaded.
In each invasion, the capillary rule is obeyed, i. e., among the few sites
surrounding the actual growth site, only the smallest empty site is invaded. 
In terms of a fluid flowing language, these supplementaries 
$N-1$ invasions would correspond to the inertia of the fluid - that is, after
the perimeter rupture has occurred, the fluid cannot stop instantaneously.
It walks $N-1$ steps further.
It may happen that this occupation of exactly $N$ sites is not 
allowed by restrictions imposed by the growth process itself and the lattice 
geometry. Many times, an incomplete walk can be found in a 'cul de sac'.
In such cases, of course, a smaller 
number of sites is in fact invaded. Any way, the next thing to do is now to
determine the new cluster's perimeter. This completes what we called {\it one
growth step}. The growth stage is then updated to $t+1$ and
the process is repeated until the cluster 
touches one of the lattice frontiers.

\section{Numerical Simulations}

As mentioned above, sometimes at a growth stage $t$, the presence of already
occupied sites prevents the walk from completing all $N$ steps, and the
walk is blocked. This means that the number of steps actually executed is
a wild function of the growth stage $t$ and, more important, it is always 
{\em smaller} or {\em equal} to $N$. On average, this would bring the mean
number of steps too far from $N$. We are then faced with the problem
of having an external (and fixed) parameter $N$ disconnected from the 
really performed mean number of steps. In order to bring the values of
these two quantities as closed as possible, we devised
the following compensation mechanism in our simulations.
Any time a walk has performed say $\bar{N}$ steps ($\bar{N} < N$) then the
debt $N - \bar{N}$ is recorded and,in the next growth stage, a total of 
$N + (N - \bar{N})$ number of steps will be permitted. So, sometimes the 
number of steps actually executed can be larger than $N$, sometimes smaller 
and the mean number of steps will be fluctuating around $N$. This scheme 
resembles those used in systems with self organized criticality. In our case, 
it is the mean number of steps which is spontaneously tuned to the value $N$.
As we shall see later, this tuning is {\em always} possible in three dimensions 
but not in two.

Let $n(t)$ be the number of steps {\it really} executed and $n_{e}(t)$ the 
{\it expected} number of steps to be executed at growth stage $t$ ($t$ is
an integer such that $t \geq 1$). The debt 
is defined as $D(t) = n_{e}(t) - n(t)$ with $n_{e}(t=1)=N$ . 
This debt should be payed in the next growth stage $t+1$ by
executing a longer walk, i. e., one with a expected number of steps 
$n_{e}(t+1)= N + D(t)$ steps. The quantities $n(t)$ and $n_{e}(t)$ are
plotted in Fig. 1 for one typical realization. From the fluctuating character 
of $n(t)$ we conclude that
the relevant parameter to be measured is the {\it mean number} of steps 
$\langle n \rangle$. If, from a total of $S$ realizations,
the number of growth steps in the $i th$ realization is $T_{i}$ then

\begin{equation}
\langle n \rangle = \frac{1}{S} \sum_{i=1}^{S} 
\frac{\sum_{t=1}^{T_{i}} n_{i}(t)}{T_{i}}
\end{equation}
where $n_{i}(t)$ stays for the number of steps really executed at the growth
stage $t$ in the $i th$ realization.

We performed numerical simulations of the N-steps invasion
percolation model on two and three dimensional lattices. For the square and 
the honeycomb lattices, sizes of $201$,$401$,$801$ and $1601$ were used and
the mean number of steps $\langle n \rangle $ (averaged over 
$400-2000$ experiments) determined for several values of $N$. Even for small
$N$, we observe the presence of blocking. In order to measure how
frequent are these blockings for fixed $N$, we calculate the quantity 

\begin{equation}
f_{b} = \frac{1}{S} \sum_{i=1}^{S} 
\frac{total \; number \; of \; blockings \; occurred \; in \; the \; i 
th \; realization} {T_{i}}
\end{equation}

As can be seen in Fig. 2 (a), this fraction of blocking 
increases with $N$ and reaches 100 \% around $N=35$ for the square lattice.
Our finite lattice size analysis, indicates that, as the thermodynamic limit
of an infinite lattice is approached, the curve extrapolates to a cusp. 
Fig. 2 (b) shows the dependence of the mean number of steps
$\langle n \rangle $ with $N$. They coincide until the upper bound 
$N_{max}=35$. This means that,
no matter how much $N$ is bigger than $35$, the ultimate N-steps invasion 
percolation model is that one with $N=35$. We conclude that the model is well 
defined only for $N \in [1,35]$ where $\langle n \rangle $ and $N$ coincide.  

This blocking phenomenon is also found in the honeycomb lattice, but
it is more frequent than that observed for the square lattice. As a result,
the breakdown is slightly smaller, i. e., $N_{max}=30$.

From studies of the kinetic growth walk model \cite{kgw}-\cite{russo} we know
that the blocking phenomenon is especially acute in two dimensions but it is 
irrelevant in higher dimensions.

For the three-dimensional case we use the simple cubic lattice with $L=51$, 
$75$,
$101$, $151$ and $201$ ($400-2000$ experiments).  Here,
the blocking is so rare that $ \langle n \rangle $ and $N$ coincide for the 
whole interval $1\leq N\leq
100$.

\section{Cluster Structure}

The clusters of ordinary invasion percolation are fractal objects in the
sense that their mean mass $\langle M \rangle $ scales with their
gyration radius $R_{g}$ with a non-integer exponent \cite{man}.  This exponent
is the fractal dimension $D_{F}$.  The values of $D_{F}$ are known from several
studies:  $D_{F}=1.89$ \cite{wil,barso,ono3} for two dimensional systems and
$D_{F}=2.52$ \cite{wil,barso} for the three dimensional case.

The fractal dimension and the mean coordination number 
\cite{ono3} are
two simple ways of characterizing the cluster's structure.
We determined these quantities for the N-steps invasion percolation model
defined on the square and simple cubic lattices.
The estimated fractal dimensions as a function of $\langle n \rangle $ are 
shown in the Table 1.  

For the square lattice only a small dependence is detected. On the other hand,
universality is definitely broken for the cubic lattice, with $D_{F}$ 
varying from $ 2.52$ to $2.77$. These results are shown in Fig. 3.

The mean coordination number $\langle z \rangle$ of the clusters is
the number of first neighbours averaged over all sites of the clusters.
For the
ordinary site invasion percolation this quantity has values:
$ 2.51(1) $ \cite{ono3} for the square lattice and 
$ 2.31(1) $ for the simple cubic lattice. Our estimated values are shown in the
Table 1.  The mean coordination number increases with $\langle
n \rangle$ but quickly stabilizes. These considerable changes of $\langle z \rangle$
with $\langle n \rangle $
indicate that some visual differences between the clusters should exist. 
This can be seen in Fig. 4. We note the formation of globules, i. e., regions
with higher densities, as a result of increasing $\langle n \rangle$.

\section{ The acceptance profile}

The acceptance profile $a(r)$ \cite{wil} concept was introduced by Wilkinson
and Willemsen to study ordinary invasion percolation.  They defined
$a(r)$ as the ratio between the number of random numbers in the interval
$[r,r+dr]$ which were accepted into the cluster to the number of random numbers
in that range which became available.  In the limit of an infinite lattice the
acceptance profile tends to a step function with the discontinuity located at
the ordinary percolation threshold $p_{c}$.

We determined the acceptance profile $a(r)$ for the N-steps invasion
percolation as a function of $\langle n \rangle $ for both the square
and simple cubic lattices.  In the Fig.5(a) we show how the finite size of the
lattice affects 
$a(r)$. As the lattice size is increased, the
acceptance profile develops a plateau up to a threshold $r_{c} \sim 0.35$
(indicating that all the small random
numbers which became available were accepted into the cluster).  
After this value, however, a tail appears and remains finite even in 
thermodynamic limit. This happens here due to the fact that the dynamic of 
the model forces invasion of some larger pores.

Fig.5(b) shows the acceptance profile for many values of $\langle
n \rangle $ for the square lattice.  The threshold $r_{c}$ of the plateau
clearly depends on the mean number of steps, i. e., $r_{c}(
\langle n \rangle)$. For $\langle n \rangle =1$ the
acceptance profile tends to a step function with $r_{c}(\langle n \rangle=1)
\simeq 0.59$ and the
ordinary invasion percolation behaviour is recovered.  Increasing
$\langle n \rangle $ diminishes $r_{c}$ until the ultimate
value $r_{c}(35) \sim 0.24$.
A similar behaviour was observed for the acceptance profile of the simple cubic
lattice, the only difference being the probable inexistence of the upper 
limited value of $N$ due the absence of blocking.

\section{Conclusions}

We proposed a new kind of invasion percolation model where the inertia of the
invader fluid was taken into account. The additional number of steps 
(or pores) $N$ governs the impetus of the fluid. In two dimensions, the
appearance of the blocking phenomenon, gives to $N$ an upper bound value 
$N_{max}$.
This means that the proposed mechanism can only be implemented with $N$
varying from $1$ to $N_{max}$.
The $N_{max}$ values were estimated to be $35$ and $30$ for square and honeycomb
lattices, respectively. There is a strong dependence between the mean 
coordination number $z$ and $N$. As $N$ is increased, we find that
some globules are formed inside the clusters structure. 
For the simple cubic lattice, the fractal dimensions 
depend strongly on $N$ and put, definetely, the N-steps invasion percolation
model in different classes of universality.

Let us now discuss in what sense we believe our model may contribute to a
better understanding of fluid flow in porous media. Inertial forces in a
porous medium are directly proportional to the square of the fluid velocity and 
inversely proportional to the pore's diameter. The Reynolds number is given
by the ratio between the inertial forces and viscous forces and, consequently,
has a linear dependence with the fluid velocity. If this number
is small (low velocity) then the Darcy's law (which states that the 
pressure gradient is linearly proportional to the velocity) can be applied. 
However, if the fluid velocity is increased but still kept below the turbulent 
regime, then a departure of Darcy's law can be experimentally measured. This 
departure is {\em macroscopically} well described by the Forchheimer equation 
\cite{bear} which adds a quadratic velocity term to the Darcy's law. It is 
believed that this quadratic term comes, essentially, from inertial forces 
contributions and it could be detected and measured by augmenting the fluid 
velocity. Recently \cite{andra}, the Forchheimer equation has been investigated 
by numerical solutions of the continuous Navier-Stokes equation. 
In our model, the parameter $N$ can be related to the fluid velocity. This
can be seen if we remember that the invaded volume by growth step (or the fluid
flux) increases with $N$. To check this directly will require to find amenable 
definitions of average pressure gradients and velocities valid in the context 
of the invasion percolation models.

\section{Acknowledgments}

We acknowledge CNPq (Conselho
Nacional de Desenvolvimento Cient\'{\i}fico e Tecnol\'ogico) and FAPESP (
Funda\c c\~ao de Amparo a Pesquisa do Estado de S\~ao Paulo ) for the financial
support.


\newpage 

\begin{center} 
Table 1 
\end{center}

\begin{center}
\begin{tabular}{||c|c|c|c|c|c||} \hline\hline
\multicolumn{3}{||c|}{Square Lattice} & \multicolumn{3}{|c||}{Simple Cubic
Lattice} \\ \hline
$\left\langle n\right\rangle $ & $D_{F}$ & $\left\langle
z\right\rangle $ & $\left\langle n\right\rangle $ & $D_{F}$ & $\left\langle
z\right\rangle $\\ \hline 
$1$ & $1.89(1)$ & $2.51(1)$ & $1$ & $2.52(2)$ & $2.31(1)$\\ \hline
$2$ & $1.89(1)$ & $2.65(1)$ & $2$ & $2.52(2)$ & $2.42(1)$\\ \hline
$4$ & $1.90(1)$ & $2.84(1)$ & $8$ & $2.58(2)$ & $2.51(1)$\\ \hline
$15$ & $1.91(1)$ & $2.89(1)$ & $15$ & $2.64(3)$ & $2.52(1)$\\ \hline
$30$ & $1.92(1)$ & $2.89(1)$ & $30$ & $2.69(3)$ & $2.52(1)$\\ \hline
$35$ & $1.92(1)$ & $2.89(1)$ & $45$ & $2.75(3)$ & $2.53(1)$\\ \hline
$40$ & $1.92(1)$ & $2.89(1)$ & $60$ & $2.77(3)$ & $2.53(1)$\\ \hline
$60$ & $1.92(1)$ & $2.89(1)$ & $100$ & $2.77(3)$ & $2.53(1)$\\ \hline\hline
\end{tabular} 
\end{center}

\newpage

\begin{center} 
FIGURE CAPTION 
\end{center}

\vspace{2.0cm}

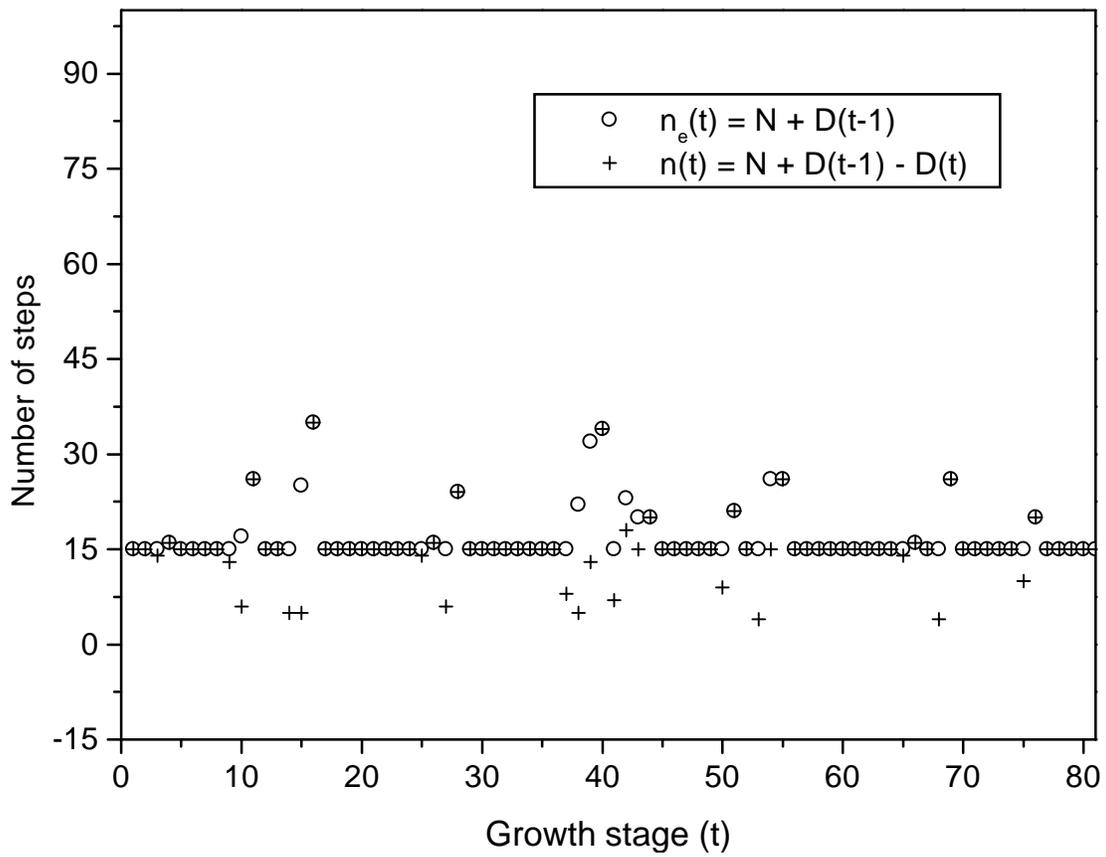
Figure 1. Simulation of the number of executed steps $n(t)$ and the number of 
expected steps $n_{e}(t)$ as a function of the growth steps $t$.

\vspace{2.0cm}

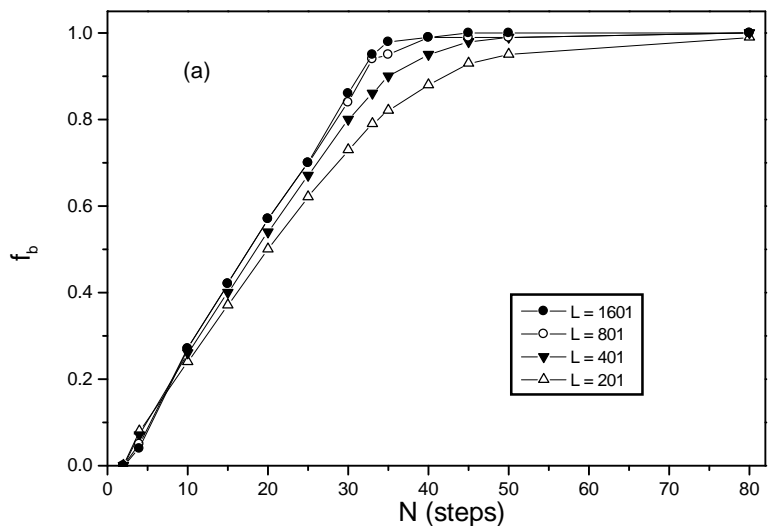
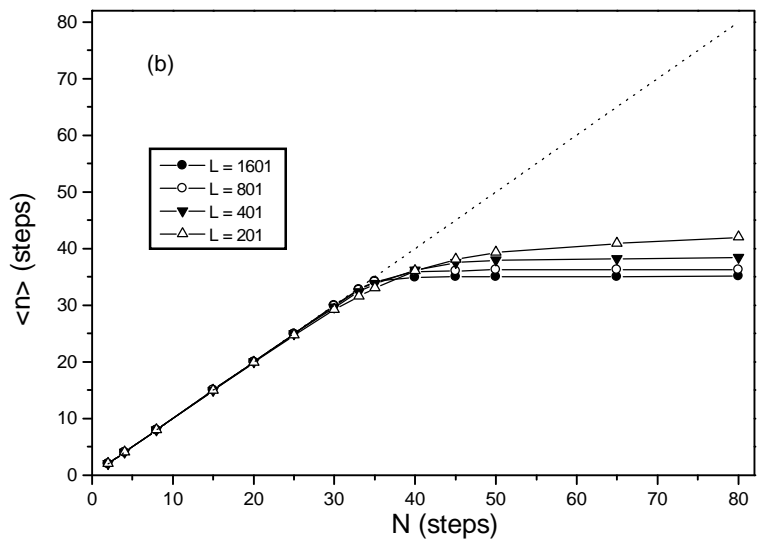
Figure 2.(a) The fraction of blocking $f_{b}$ as a function of $N$ for
the square lattice, (b) The mean number of steps versus
$N$. The dotted line is the straight $\langle n \rangle$=$N$.

\vspace{2.0cm}

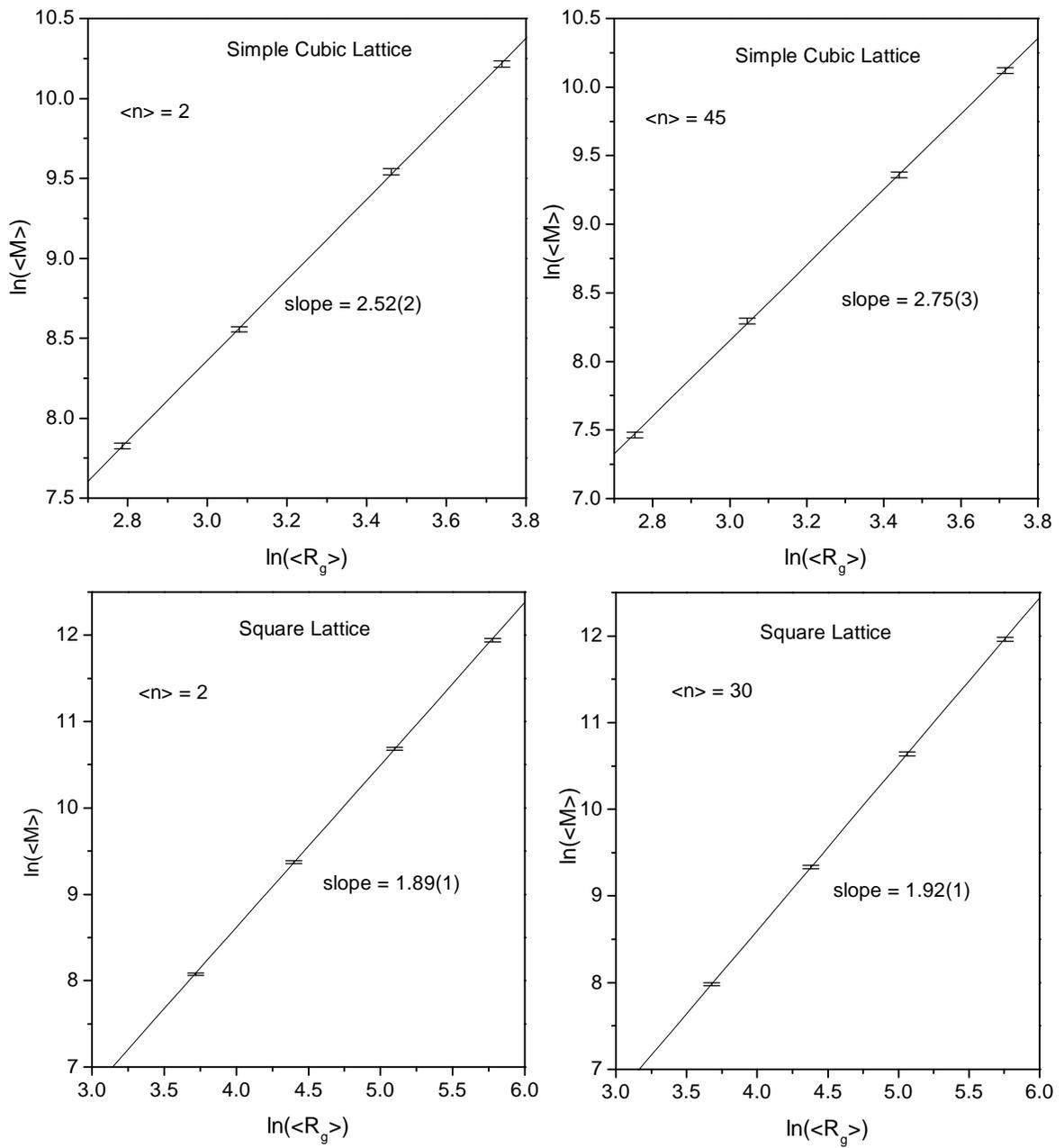
Figure 3. Logarithm of the cluster average mass plotted versus the 
logarithm of the average gyration radius for the cubic and square lattices
and for some values of $\langle n \rangle$.

Figure 4. The upper figure shows a typical cluster of the ordinary invasion 
percolation model for a square lattice with $L=201$. The lower is a 
cluster of the N-steps invasion percolation model with $\langle n \rangle = 30$
and $L=201$. 

\vspace{2.0cm}

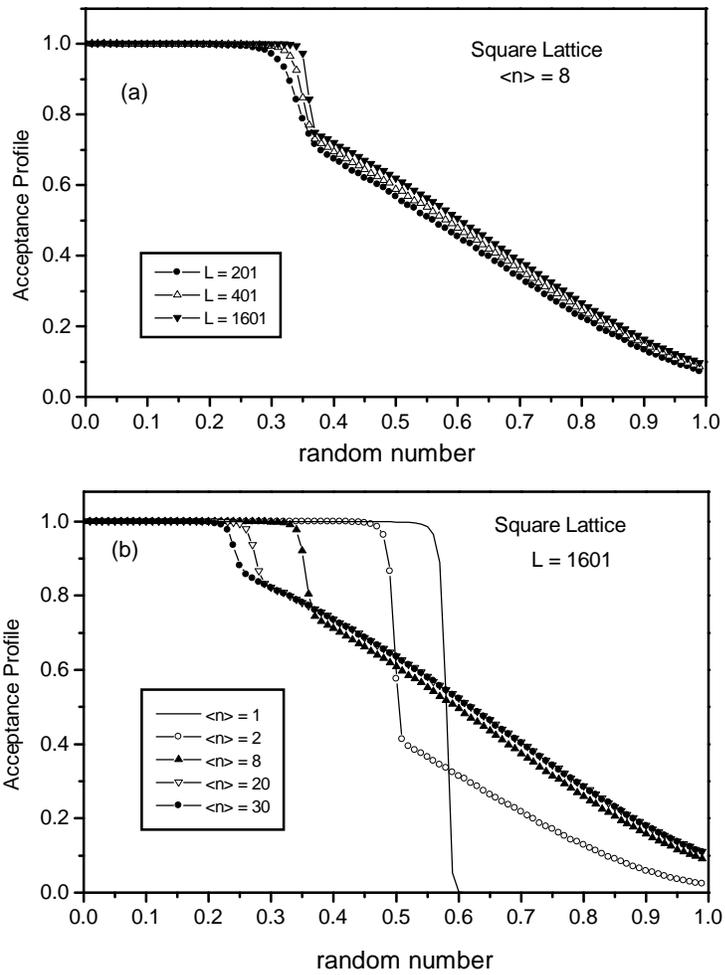
Figure 5. The dependence of the acceptance profiles with the random number 
$r$ as the
lattice size $L$ and the mean number of steps $\langle n \rangle $ are changed
( (a) and (b), respectively).

\newpage

\begin{center} 
TABLE CAPTION 
\end{center}

\vspace{2.0cm}

Table 1.  The fractal dimensions and mean coordination numbers of the clusters
on the square and simple cubic lattices for several values of
$\langle n \rangle $.

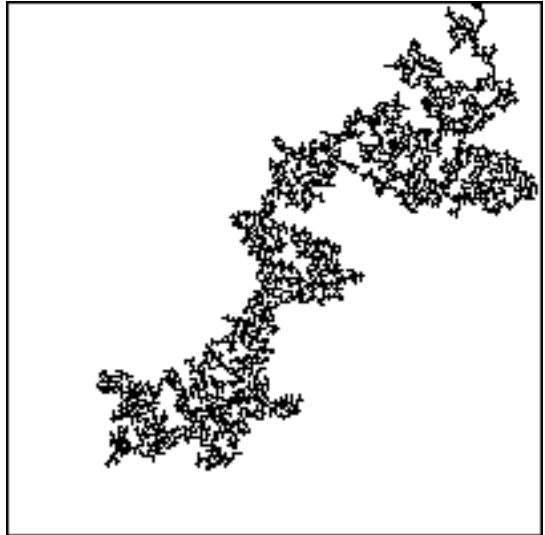
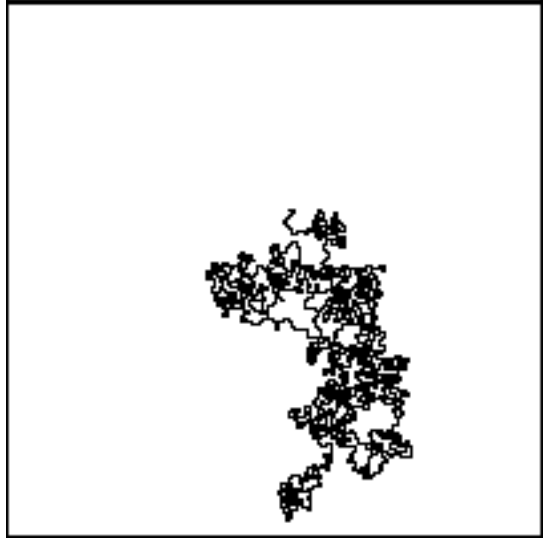
\end{document}